# Hemodynamic Overlap Between Ruptured and Unruptured Cerebral Aneurysms Using an In-House Parallel C++ Finite Element Solver

Reza Bozorgpour *[1], Mohammadreza Soltany Sadrabadi[2]

[1]Department of Biomedical Engineering, University of Wisconsin-Milwaukee, Milwaukee, WI, USA
[2]Department of Mechanical Engineering, Northern Arizona University, Flagstaff, AZ, USA

*Corresponding author

E-mail: bozorgp2@uwm.edu

## Abstract

Cerebral aneurysm rupture has long been associated with abnormal hemodynamic conditions, particularly wall shear stress (WSS)-related flow behavior. Although aneurysm rupture is widely recognized as a multifactorial phenomenon involving complex interactions between geometry, flow structures, and multiple hemodynamic quantities, WSS-based parameters remain among the most commonly reported metrics in computational aneurysm studies. In the present study, computational fluid dynamics simulations were performed using an in-house parallel C++ finite element solver to investigate velocity streamlines, wall shear stress (WSS), and time-averaged wall shear stress (TAWSS) in patient-specific ruptured and unruptured cerebral aneurysm models. The computed flow fields demonstrated substantial hemodynamic overlap between the two groups, with elevated WSS and TAWSS regions observed in both ruptured and unruptured aneurysms. Similar high-velocity inflow structures and localized shear concentration patterns were identified across different cases, demonstrating that elevated shear-related magnitudes can be present in aneurysms with different rupture statuses. The results show that comparable WSS- and TAWSS-based hemodynamic characteristics may exist in aneurysms with different clinical outcomes, emphasizing the complexity and case-dependent nature of cerebral aneurysm hemodynamics. Rather than proposing rupture prediction criteria, the present study provides a focused computational assessment of hemodynamic similarities between ruptured and unruptured aneurysms and highlights the importance of careful interpretation of isolated WSS- and TAWSS-based analyses in rupture assessment.

## 1. Introduction

Intracranial aneurysms represent pathological enlargements of cerebral arteries that pose a serious clinical risk due to their potential to rupture. Although they are relatively common, affecting up to 6% of the population, the proportion that rupture is small; however, when rupture occurs, it often leads to subarachnoid hemorrhage with a mortality rate approaching 60% [1-4]. The mechanisms underlying aneurysm formation and progression are complex and multifactorial. A range of contributing factors, including genetic predisposition, hypertension, infections, smoking, alcohol use, and aging, have been implicated in altering both vascular integrity and blood flow characteristics [5, 6]. These alterations promote structural weakening of the arterial wall and disrupt normal hemodynamic conditions, creating an environment conducive to aneurysm development. Additional processes such as inflammation and flow-induced stresses further influence whether an aneurysm remains stable or progresses toward rupture [7-11]. Despite extensive investigation, the combined effects of these biological and mechanical factors are not yet fully resolved. Therefore, understanding the coupled behavior of vessel wall mechanics and blood flow is critical for identifying regions that are more vulnerable to rupture. This clinical burden is further reflected at the population level. Intracranial aneurysms are present in approximately 3.2% of individuals [12], and in the United States alone, rupture leads to nearly 30,000 cases each year, contributing to 0.4–0.6% of overall mortality [13]. Even when not immediately fatal, outcomes remain poor: the one-year mortality rate reaches 35.3%, and among survivors, only 48.7% regain baseline function, while 16% experience persistent disability, underscoring the long-term impact of the disease [14, 15].

Recent advances in medical imaging and computational resources have made it possible to reconstruct patient-specific cerebrovascular geometries and analyze blood flow patterns with high fidelity. Within this framework, computational fluid dynamics (CFD) has emerged as a valuable approach for investigating the underlying mechanisms of aneurysm behavior and translating complex flow phenomena into clinically meaningful insights [16]. In parallel, growing evidence highlights the importance of local hemodynamic conditions in driving aneurysm initiation and progression, suggesting that flow-related forces play a central role beyond traditional risk indicators [17]. While clinical factors such as smoking, hypertension, and connective tissue disorders, along with aneurysm features including size, location, and morphology, are known to influence rupture risk, these alone do not fully explain the observed variability in outcomes [18-20].

The study of blood flow in intracranial aneurysms increasingly relies on computational approaches that combine physical laws of fluid motion with numerical algorithms and modern computing [21]. Through this lens, hemodynamic quantities can be extracted and used to interpret mechanisms underlying aneurysm progression and rupture. In particular, forces acting along the vessel wall have been widely investigated, as both elevated and diminished shear environments are thought to disrupt endothelial function, promote inflammatory activity, and contribute to structural degradation of the arterial wall [22, 23].

Early computational studies adopted simplified assumptions, typically treating the vessel wall as rigid and the flow as time-invariant, primarily to reduce computational demands and based on their applicability in larger vascular segments [24]. As computational resources and imaging techniques have advanced, these simplifications have gradually been relaxed. More recent efforts incorporate pulsatile flow conditions, enabling the evaluation of time-dependent indices such as oscillatory shear index and averaged shear metrics, while further developments include fluid-structure interaction models that capture the coupling between blood flow and wall deformation [25-29]. Although not yet standard practice, these approaches provide a more realistic representation of physiological conditions and improve the predictive capability of simulations.

Despite extensive efforts in computational hemodynamics, inconsistencies in modeling approaches and numerical implementations continue to limit the reproducibility and clinical translation of existing findings. To address these challenges, the present study develops an in-house parallel finite element framework for simulating blood flow in patient-specific cerebral aneurysm geometries. The solver is designed to efficiently handle large-scale vascular models while accurately resolving flow fields and associated hemodynamic quantities. Using this computational platform, key parameters such as wall shear stress, oscillatory shear index, and related metrics are systematically evaluated to investigate their relationship with aneurysm rupture status. By combining high-fidelity numerical modeling with a consistent computational framework, this work aims to provide clearer insight into the role of hemodynamics in aneurysm rupture and contribute toward more reliable predictive assessment.

## 2. Method

A computational framework based on the finite element method was developed to simulate blood flow in patient-specific cerebral aneurysm geometries and investigate the relationship between hemodynamic variables and rupture status. The governing incompressible Navier-Stokes equations were solved under physiologically realistic pulsatile flow conditions. High-resolution vascular models were discretized using unstructured meshes, and simulations were performed using a highly parallelized implementation employing domain decomposition and efficient sparse matrix storage. Key hemodynamic parameters, including Velocity profile, WSS and TAWSS were computed over the cardiac cycle and analyzed to identify differences between ruptured and unruptured aneurysms.

### 2.1. Geometry and Data Source

Patient-specific cerebral aneurysm geometries were obtained from the publicly available Aneurisk repository [30]. The dataset consists of three-dimensional reconstructions of intracranial arterial segments with associated rupture status. A subset of aneurysms including both ruptured and unruptured cases was selected for the present analysis. The geometries were preprocessed to ensure numerical stability and suitability for computational analysis. This process included removal of imaging artifacts, smoothing of surface irregularities, and truncation or extension of inlet and outlet branches to promote the development of physiologically meaningful flow fields. All preprocessing steps were performed in a consistent manner across all models to ensure uniform treatment of the dataset. It is acknowledged that geometry preprocessing and numerical implementation may introduce variability across different computational frameworks and user workflows, potentially influencing absolute hemodynamic quantities. To minimize such effects within this study, a standardized preprocessing protocol was applied to all cases. Accordingly, the results are interpreted in a comparative framework, with emphasis placed on relative differences between ruptured and unruptured aneurysms rather than absolute magnitudes. Such variability is inherent to patient-specific computational hemodynamics and has been widely reported in the literature. Following preprocessing, all geometries were verified to be watertight and suitable for volumetric meshing and subsequently discretized into finite element meshes while preserving the anatomical features relevant to local hemodynamic behavior.

## 2.2. Blood Properties and Flow Assumptions

Blood was modeled as an incompressible, Newtonian fluid, which is a commonly adopted assumption for flow in large cerebral arteries. The fluid density and dynamic viscosity were taken as $\rho = 1060$ kg/m$^3$and $\mu = 0.0035$ Pa. s, respectively. Vessel walls were assumed to be rigid, and a no-slip boundary condition was imposed at the lumen surface. Flow within the vascular models was assumed to be laminar due to the relatively low Reynolds numbers in intracranial circulation. A time-dependent pulsatile inflow condition was prescribed to capture the transient characteristics of physiological blood flow. The flow field was governed by the three-dimensional, unsteady, incompressible Navier–Stokes equations. These consist of the continuity equation, enforcing mass conservation, and the momentum equation, describing the balance of inertial, pressure, and viscous forces:

$$\nabla \cdot \mathbf{u} = 0$$

$$\rho \left( \frac{\partial \mathbf{u}}{\partial t} + \mathbf{u} \cdot \nabla \mathbf{u} \right) = -\nabla p + \mu \nabla^2 \mathbf{u} \quad \text{Eq. 1}$$

where $\mathbf{u}$ is the velocity vector and $p$ is the pressure field.

## 2.3. Mesh Generation and Independence Study

The aneurysm geometries were preprocessed prior to meshing using a combination of Blender, Meshmixer, and MeshLab. These tools were employed for surface repair, including removal of artifacts, smoothing of irregularities, and ensuring watertight geometries suitable for volumetric meshing. All preprocessing steps were applied consistently across the dataset to maintain uniformity. Following preprocessing, the geometries were discretized into unstructured tetrahedral meshes using Gmsh [31]. The mesh generation process employed Delaunay-based tetrahedral meshing to enable robust handling of complex vascular topologies. Local mesh refinement was applied in regions of high curvature and within aneurysm sacs to ensure accurate resolution of flow features. In addition, prism-based boundary layer meshes were generated near vessel walls to adequately resolve near-wall velocity gradients, which are critical for accurate computation of WSS and TAWSS. Different mesh densities were prescribed in selected regions to achieve an appropriate balance between numerical accuracy and computational efficiency.

A mesh independence study was performed to ensure that the computed hemodynamic quantities were not sensitive to spatial discretization. Multiple meshes with increasing element densities were generated for representative cases, and key parameters, including WSS and OSI, were evaluated. Mesh convergence was assumed when further refinement resulted in negligible variation in these quantities. Based on this analysis, an optimal mesh density was selected to balance computational accuracy and efficiency.

## 2.4. Boundary Conditions

At each outlet, the downstream vascular effect was approximated using a resistance-based boundary condition derived from laminar flow theory. The relationship between pressure drop and flow rate across each outlet was defined as:

$$\Delta p = QR \quad \text{Eq. 2}$$

where $Q$ denotes the volumetric flow rate and $R$ represents the effective hydraulic resistance. The resistance values were estimated using:

$$R = \frac{8\mu L}{\pi a^4} \qquad \text{Eq. 3}$$

with $\mu$ being the dynamic viscosity, $L$ the characteristic vessel length, and $a$ the outlet radius. This approach provides an approximate distribution of flow among the outlets in the absence of patient-specific downstream vascular data. Vessel walls were modeled as rigid with a no-slip condition imposed at the fluid-wall interface. Simulations were advanced over multiple cardiac cycles until periodic flow behavior was achieved, after which hemodynamic quantities were evaluated.

## 2.5. Temporal Discretization and Simulation Setup

The governing equations were solved in a time-dependent framework to capture the pulsatile nature of blood flow. A uniform time step was employed to discretize the cardiac cycle, ensuring adequate temporal resolution of transient flow features. The time step size was selected to balance numerical stability and computational efficiency while resolving rapid variations in velocity and wall shear stress. Simulations were performed over multiple cardiac cycles to eliminate the influence of initial transients. Hemodynamic quantities were extracted only after periodic flow behavior was achieved, ensuring that the solution was independent of the initial conditions. The duration of the cardiac cycle was set to match the prescribed inlet waveform, and sufficient temporal sampling was used to accurately compute cycle-averaged parameters. All simulations were initialized from a quiescent state with zero velocity and pressure fields. Convergence within each step was achieved through iterative solution of the discretized equations, and the overall simulation was considered stable once successive cycles exhibited negligible variation in key flow variables.

## 2.6. Numerical Framework

### 2.6.1. Finite Element Formulation

The governing incompressible Navier–Stokes equations were discretized in space using the Galerkin finite element method. The velocity and pressure fields were approximated using interpolation (shape) functions such as:

$$\mathbf{u} \approx \sum_{i=1}^{n} N_i\, \mathbf{u}_i, \qquad p \approx \sum_{i=1}^{n} N_i\, p_i \qquad \text{Eq. 4}$$

where $N_i$ are the shape functions and $\mathbf{u}_i$, $p_i$ are the nodal values of velocity and pressure, respectively.

Multiplying the momentum equation by a test function $\omega$ and integrating over the domain $\Omega$:

$$\int_\Omega \omega \cdot \rho \left( \frac{\partial \mathbf{u}}{\partial t} + \mathbf{u} \cdot \nabla \mathbf{u} \right) d\Omega = \int_\Omega \omega \cdot (-\nabla p + \mu\, \nabla^2 \mathbf{u}) d\Omega \qquad \text{Eq. 5}$$

After applying integration by parts to the viscous term:

$$\int_{\Omega} \omega \cdot \rho \frac{\partial \mathbf{u}}{\partial t} d\Omega + \int_{\Omega} \omega \cdot \rho(\mathbf{u} \cdot \nabla \mathbf{u}) d\Omega + \int_{\Omega} \nabla \omega : \mu \nabla \mathbf{u} \, d\Omega - \int_{\Omega} (\nabla \cdot \omega) p \, d\Omega = \int_{\Gamma} \omega \cdot \mathbf{t} \, d\Gamma \quad \text{Eq. 6}$$

Substituting the shape functions and assembling over all elements leads to the following system:

**Mass matrix**

$$M_{ij} = \int_{\Omega} \rho \, N_i N_j \, d\Omega \quad \text{Eq. 7}$$

**Convection matrix**

$$C_{ij} = \int_{\Omega} \rho \, N_i (\mathbf{u} \cdot \nabla N_j) \, d\Omega \quad \text{Eq. 8}$$

**Diffusion (stiffness) matrix**

$$K_{ij} = \int_{\Omega} \mu \nabla N_i \cdot \nabla N_j \, d\Omega \quad \text{Eq. 9}$$

**Pressure-velocity coupling matrix**

$$G_{ij} = \int_{\Omega} (\nabla \cdot \mathbf{N}_i) N_j \, d\Omega \quad \text{Eq. 10}$$

The resulting system of equations can be written in block matrix form as:

$$\begin{bmatrix} M + C + K & -G \\ G^T & 0 \end{bmatrix} \begin{bmatrix} \mathbf{u} \\ p \end{bmatrix} = \begin{bmatrix} \mathbf{f} \\ 0 \end{bmatrix} \quad \text{Eq. 11}$$

#### 2.6.2. Domain Decomposition (METIS)

Domain decomposition was performed using the METIS library [32, 33] to enable efficient parallel computation of the governing equations. METIS provides advanced graph-based partitioning algorithms that decompose the computational mesh into subdomains while preserving element connectivity and minimizing inter-partition communication.

In the present work, the finite element mesh was represented as a graph in which elements correspond to nodes and their adjacency defines the connectivity. METIS was employed to partition this graph into a set of balanced subdomains, ensuring an approximately equal distribution of computational workload across processors. At the same time, the partitioning process minimizes the number of interface elements between subdomains, thereby reducing communication overhead during the assembly and solution phases.

The use of METIS significantly improves the scalability and efficiency of the solver, particularly for large-scale simulations involving complex vascular geometries. By reducing communication costs and maintaining load balance, the domain decomposition strategy enables effective parallelization of the finite element computations.

#### 2.6.3. Matrix Assembly and Storage (CSR)

To illustrate the construction of the CSR data structure and its integration with domain decomposition, the mesh shown in Fig. 1 is considered as a representative example. The global sparsity pattern of the finite element system is obtained directly from the element connectivity $e_1 = (1,2,5)$, $e_2 = (2,3,5)$, $e_3 = (1,5,4)$, and $e_4 = (4,5,6)$. The resulting global matrix $A \in \mathbb{R}^{6\times 6}$is sparse, with nonzero entries corresponding to neighboring nodes. For example, node 1 is connected to nodes $\{1, 2, 4, 5\}$, while node 5 is connected to all nodes. The matrix is stored in compressed sparse row (CSR) using three arrays including $row_{ptr}$, $col_{idx}$ and $val$ such that

$$A(i,j) = \text{val}[k], \qquad j = \text{col_idx}[k], \qquad k \in [\text{row_ptr}[i], \text{row_ptr}[i+1]) \qquad \text{Eq. 12}$$

The domain is partitioned into two subdomains corresponding to the MPI processors shown in Figure 1. Processor 0 contains elements $\{e_1, e_2\}$with node set $\{1, 2, 3, 5\}$, while Processor 1 contains elements $\{e_3, e_4\}$with node set $\{1, 4, 5, 6\}$. For each processor $p$, a local-to-global mapping is defined as

$$\text{L2G}_p = \{n_0, n_1, \ldots, n_{n_p-1}\},$$

which maps local indices to global node numbers.

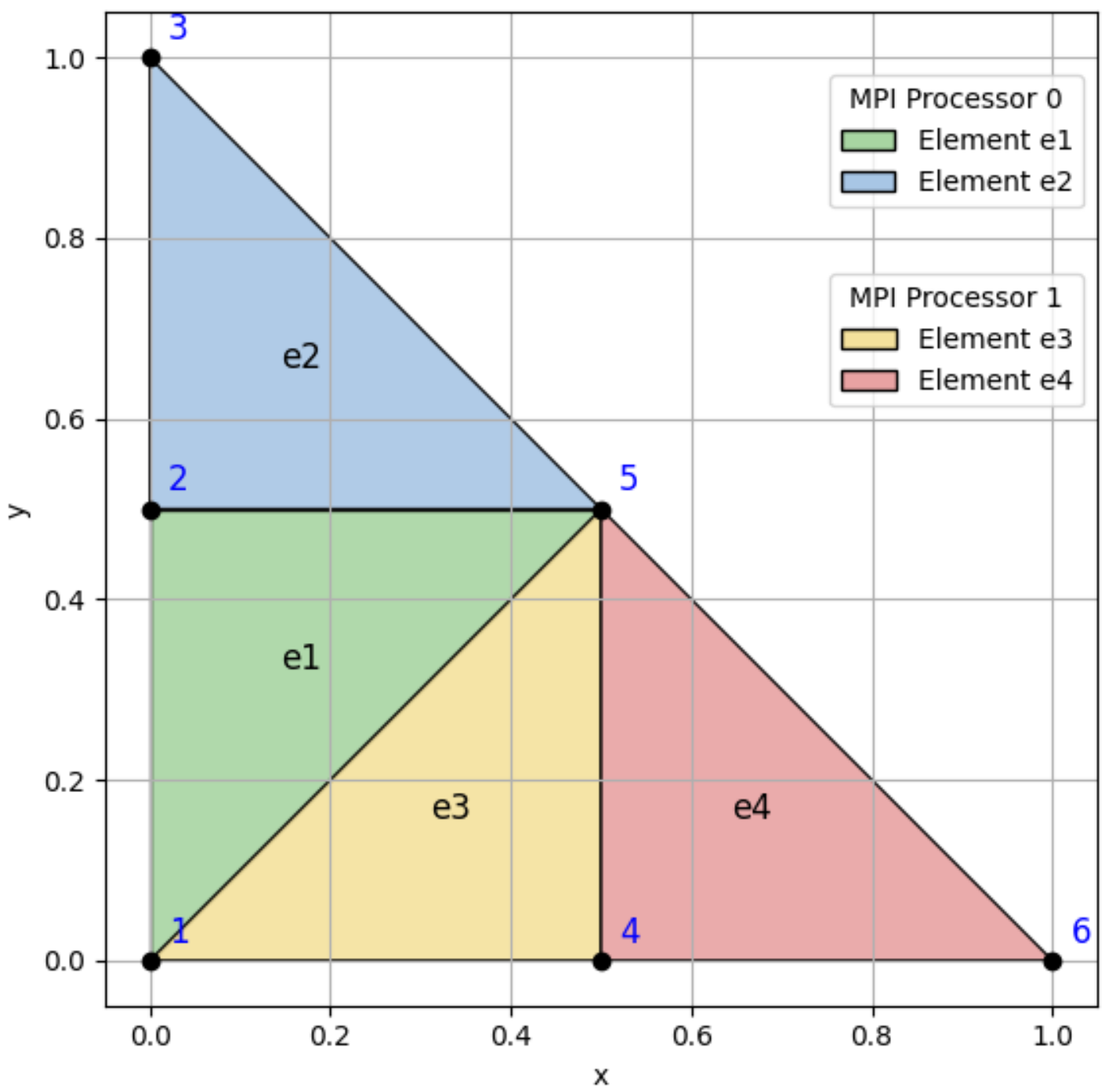

Figure 1: Triangular mesh with node and element numbering, partitioned into two MPI subdomains.

The local matrices are assembled in CSR format using local indexing, and global consistency is maintained through this mapping. Shared nodes (e.g., nodes 1 and 5) appear in multiple subdomains and require communication during the parallel solution process.

## 2.7. Solver

The linear systems arising at each time step were solved using an iterative Krylov subspace method applied directly to the sparse matrix stored in CSR format. This ensures that all matrix-vector operations are performed using only the nonzero entries of the system, thereby preserving memory efficiency and reducing computational costs. The matrix-vector product, which forms the core operation of the iterative solver, was implemented using the CSR structure through row-wise traversal of the nonzero entries. Specifically, for each row $i$, the operation is evaluated as

$$Ax)_i = \sum_{k=\text{row_ptr}[i]}^{\text{row_ptr}[i+1]-1} \text{val}\,[k] \cdot x_{\text{col_idx}[k]}$$

This formulation avoids explicit storage of zero entries and enables efficient sparse computation. The global system was distributed across MPI processes according to the domain decomposition. Each processor performed local matrix-vector operations on its subdomain using the local CSR structure, while values corresponding to shared (interface) nodes were updated through inter-processor communication at each iteration. This ensures consistency of the solution across subdomain boundaries. An iterative solver based on the GMRES method was employed to solve the system $Ax = b$, as it is well-suited for large, sparse, and non-symmetric systems arising from the discretized Navier-Stokes equations. Convergence was monitored using the relative residual norm, and the iterations were terminated once a tolerance of $10^{-6}$ was achieved.

## 2.8. Hemodynamic Parameter Calculation

### 2.8.1. Wall Shear Stress (WSS)

Wall shear stress is the tangential component of the viscous stress tensor at the wall:

$$\boldsymbol{\tau}_w = \boldsymbol{\sigma} \cdot \mathbf{n} - (\mathbf{n} \cdot \boldsymbol{\sigma} \cdot \mathbf{n})\mathbf{n} \qquad \text{Eq. 13}$$

where $\boldsymbol{\tau}_w$, $\boldsymbol{\sigma}$, $\mathbf{n}$ are wall shear stress vector, Cauchy stress tensor and unit normal vector at the wall, respectively. For Newtonian incompressible flow

$$\boldsymbol{\tau}_w = \mu[\nabla\mathbf{u} + (\nabla\mathbf{u})^T] \cdot \mathbf{n} - (\mathbf{n} \cdot \mu[\nabla\mathbf{u} + (\nabla\mathbf{u})^T] \cdot \mathbf{n})\mathbf{n} \qquad \text{Eq. 14}$$

The magnitude of the wall shear stress is then defined as:

$$\text{WSS} = | \boldsymbol{\tau}_w | \qquad \text{Eq. 15}$$

which represents the instantaneous shear force per unit area acting tangentially on the vessel wall.

### 2.8.2. Time-Averaged Wall Shear Stress (TAWSS)

TAWSS is defined as the temporal average of the WSS magnitude over one cardiac cycle:

$$\text{TAWSS} = \frac{1}{T}\int_0^T | \boldsymbol{\tau}_w(t) | \, dt \qquad \text{Eq. 16}$$

where $T$denotes the duration of the cardiac cycle and $\boldsymbol{\tau}_w(t)$is the instantaneous wall shear stress vector. TAWSS characterizes the cumulative shear exposure experienced by the vessel wall over time and is commonly used to assess long-term hemodynamic effects on endothelial cells. Unlike instantaneous WSS, TAWSS does not account for directional variations in shear stress but instead reflects the overall magnitude of shear forces acting throughout the cardiac cycle.

# 3. Results

A total of eight patient-specific cerebral aneurysm models obtained from the AneuRisk database were investigated in the present study. To provide a balanced comparison between rupture states while minimizing potential demographic variability, only female cases were considered. The dataset consisted of four unruptured aneurysms from patients aged 53, 51, 35, and 42 years, and four ruptured aneurysms from patients aged 39, 33, 49, and 36 years. CFD simulations were performed for all cases using the proposed in-house parallel C++ finite element solver, and the resulting velocity streamlines, WSS, and TAWSS distributions were comparatively analyzed between the two groups. A mesh independence study was performed for each aneurysm model to ensure that the computed hemodynamic quantities were not significantly influenced by mesh resolution. Successive mesh refinements were carried out until negligible variations were observed in the evaluated flow characteristics. The final meshes used for the simulations were selected to achieve an appropriate balance between numerical accuracy and computational efficiency. Table X summarizes the mesh characteristics for all investigated cases, including the total number of elements and the corresponding mesh quality metrics.

Table 1: Mesh Characteristics and Quality Metrics for the Investigated Aneurysm Models

| Patient ID | Age | Rapture Status | Number of Element | Average Mesh quality |
|---|---|---|---|---|
| C0001 | 53 | Unruptured | 3144384 | 0.6722 |
| C0020 | 51 | Unruptured | 6636362 | 0.6642 |
| C0024 | 35 | Unruptured | 2704998 | 0.663 |
| C0026 | 42 | Unruptured | 4084453 | 0.6719 |
| C0009 | 39 | Raptured | 9452395 | 0.6914 |
| C0068 | 33 | Raptured | 5078633 | 0.6848 |
| C0073 | 49 | Raptured | 3952159 | 0.6636 |
| C0094 | 36 | Raptured | 5057361 | 0.6645 |

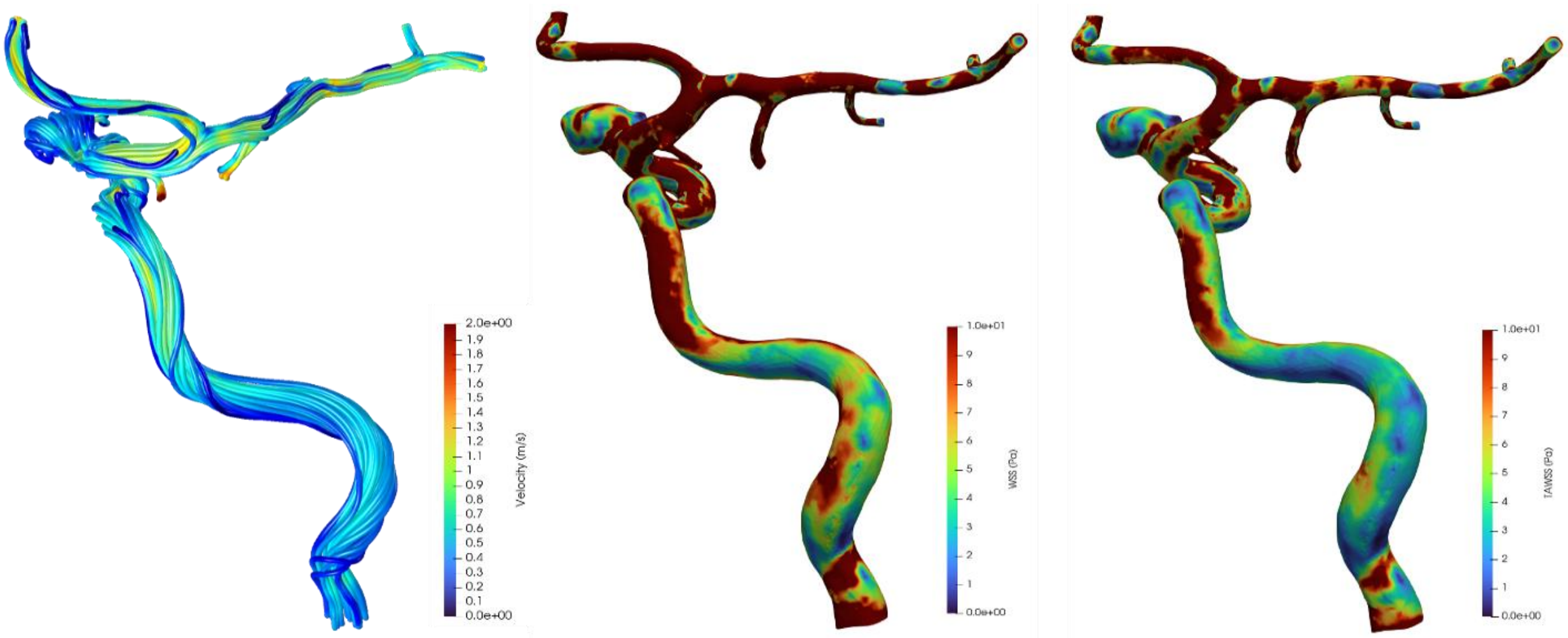

Figure 2: Velocity, WSS, and TAWSS distributions for patient C0001 (female, 53 years old, unruptured aneurysm).

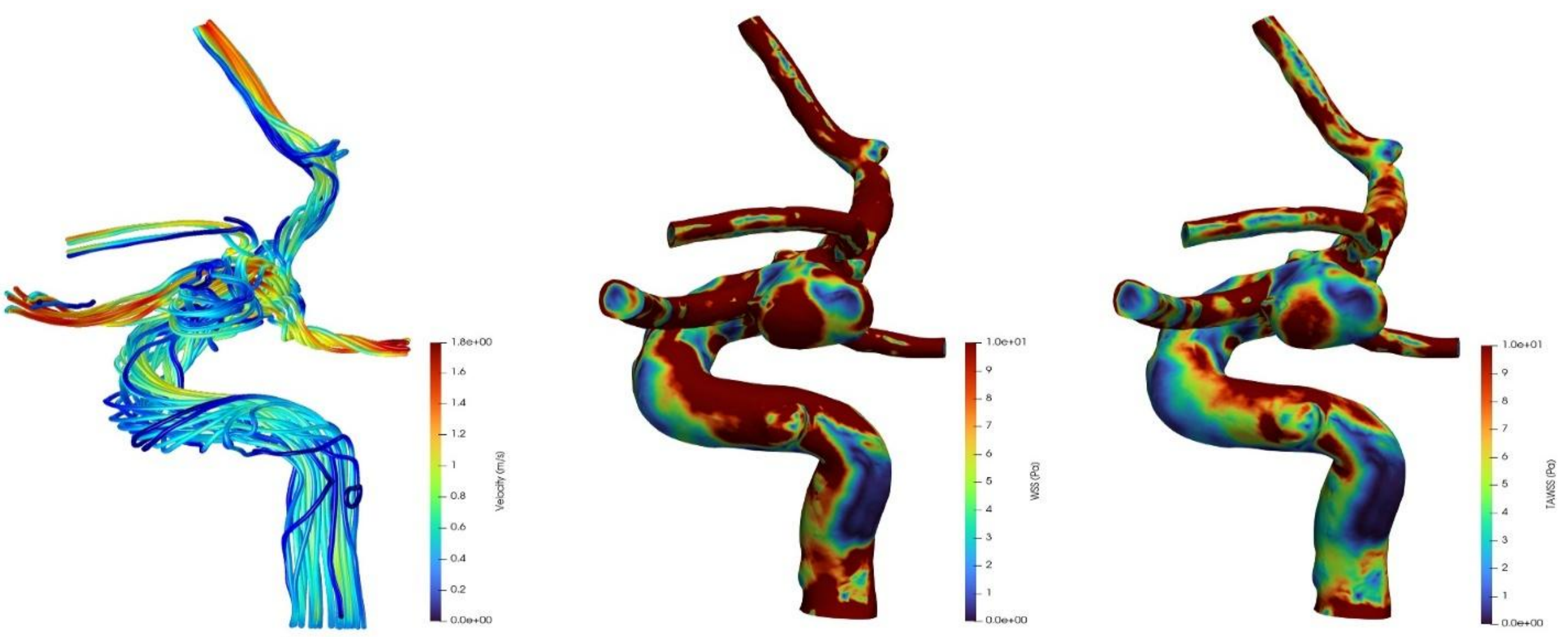

Figure 3: Velocity, WSS, and TAWSS distributions for patient C0020 (female, 51 years old, unruptured aneurysm).

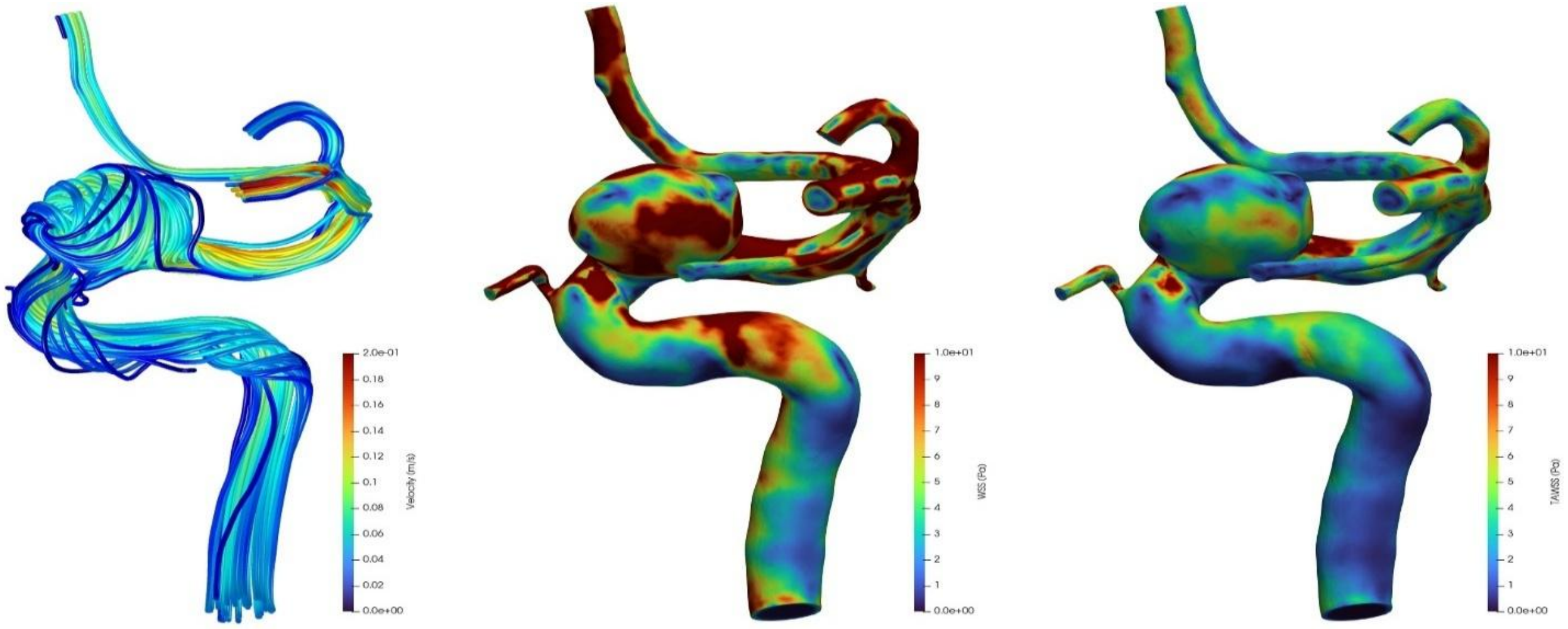

Figure 4: Velocity, WSS, and TAWSS distributions for patient C0024 (female, 35 years old, unruptured aneurysm).

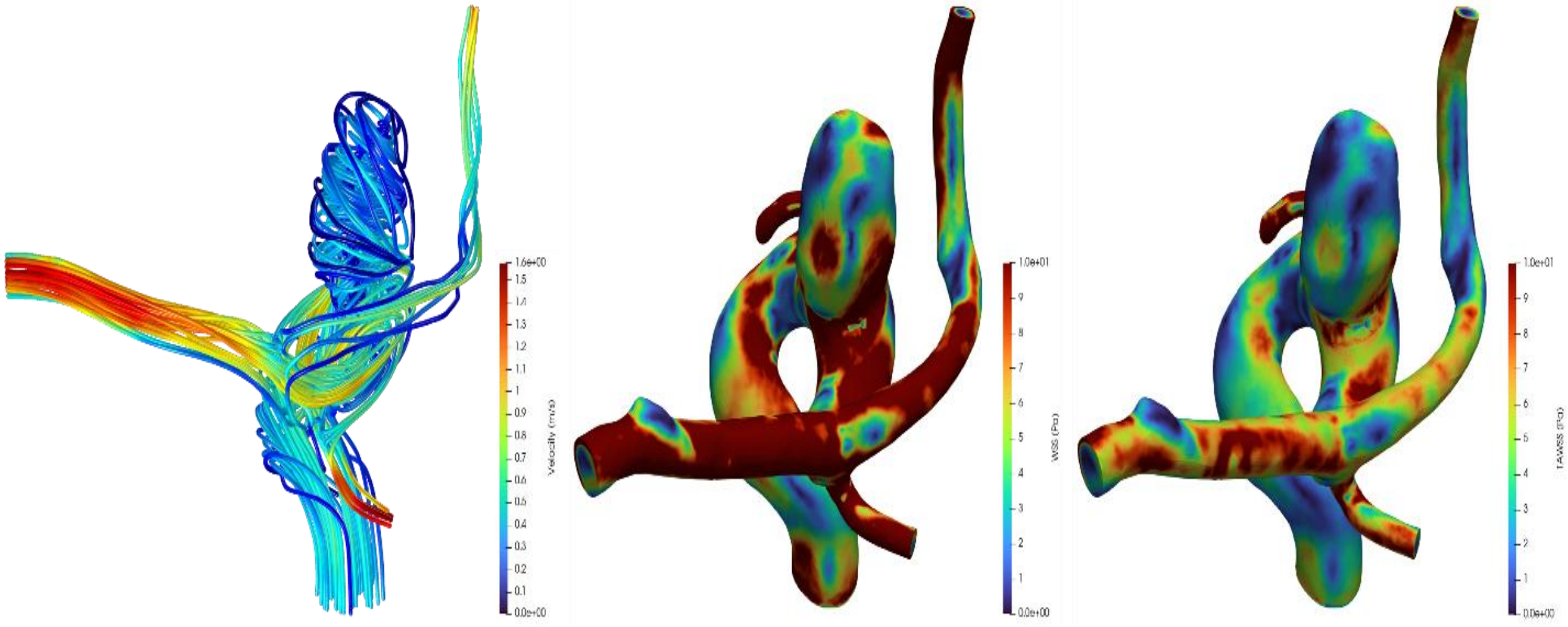

Figure 5: Velocity, WSS, and TAWSS distributions for patient C0026 (female, 42 years old, unruptured aneurysm).

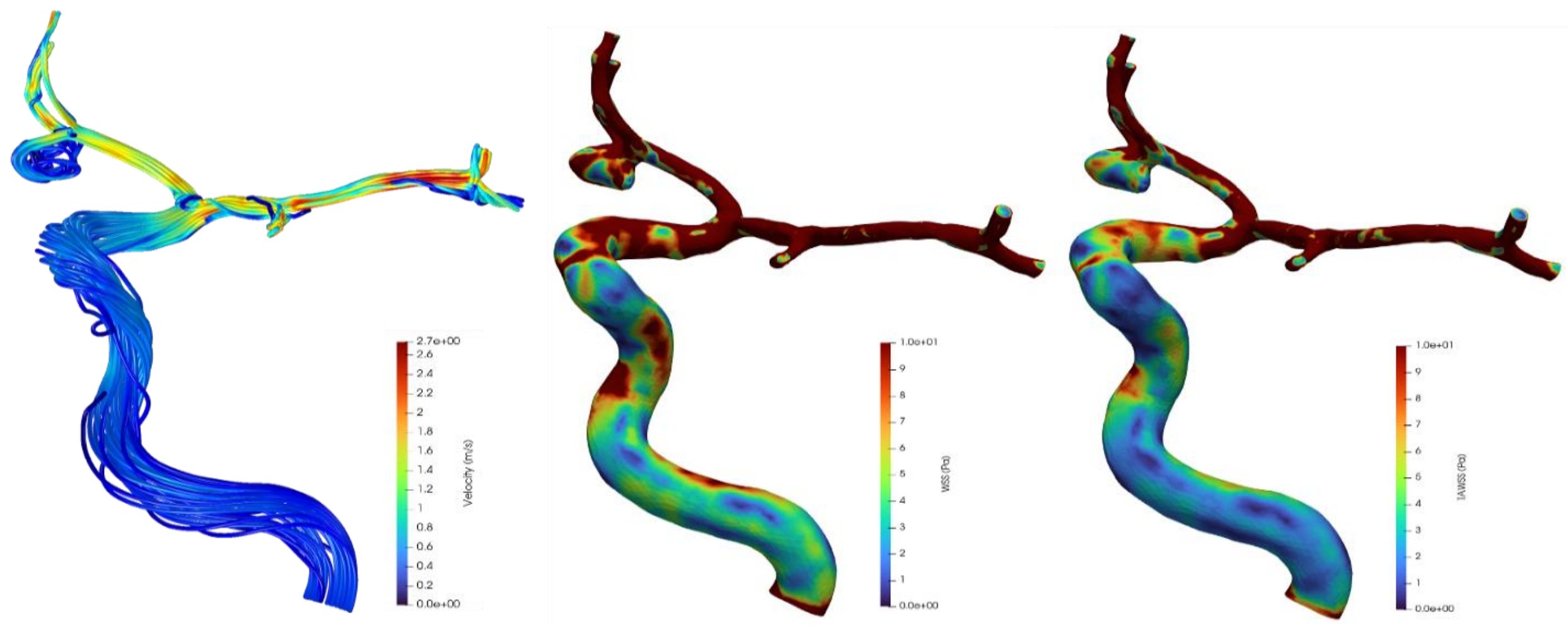

Figure 6: Velocity, WSS, and TAWSS distributions for patient C0009 (female, 39 years old, ruptured aneurysm).

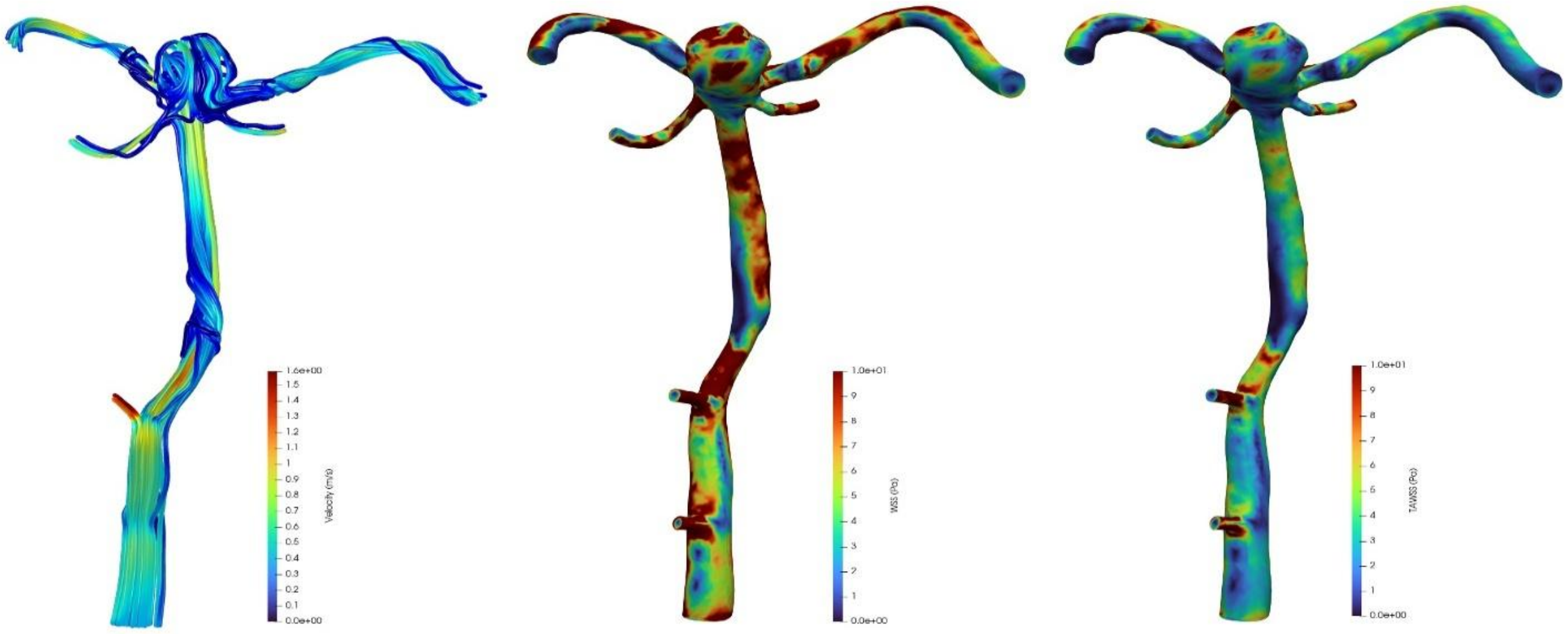

Figure 7: Velocity, WSS, and TAWSS distributions for patient C0068 (female, 33 years old, ruptured aneurysm).

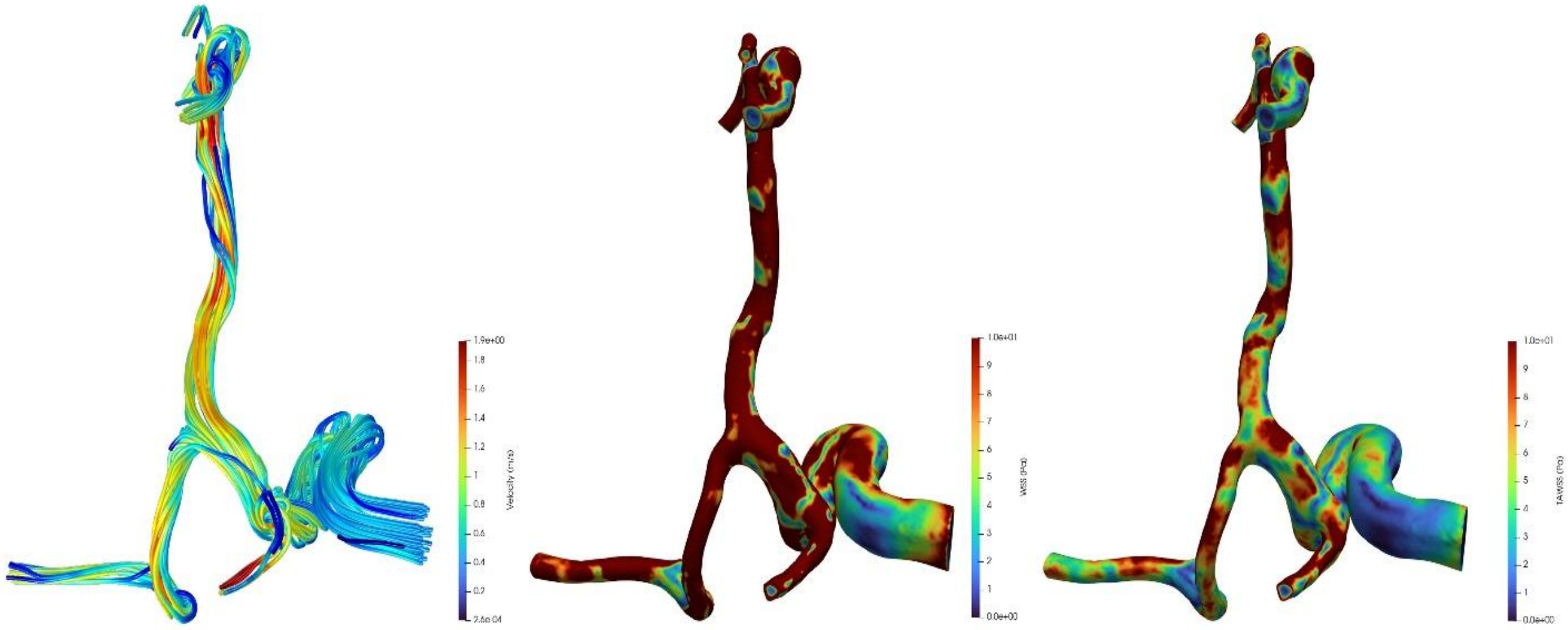

Figure 8: Velocity, WSS, and TAWSS distributions for patient C0073 (female, 49 years old, ruptured aneurysm).

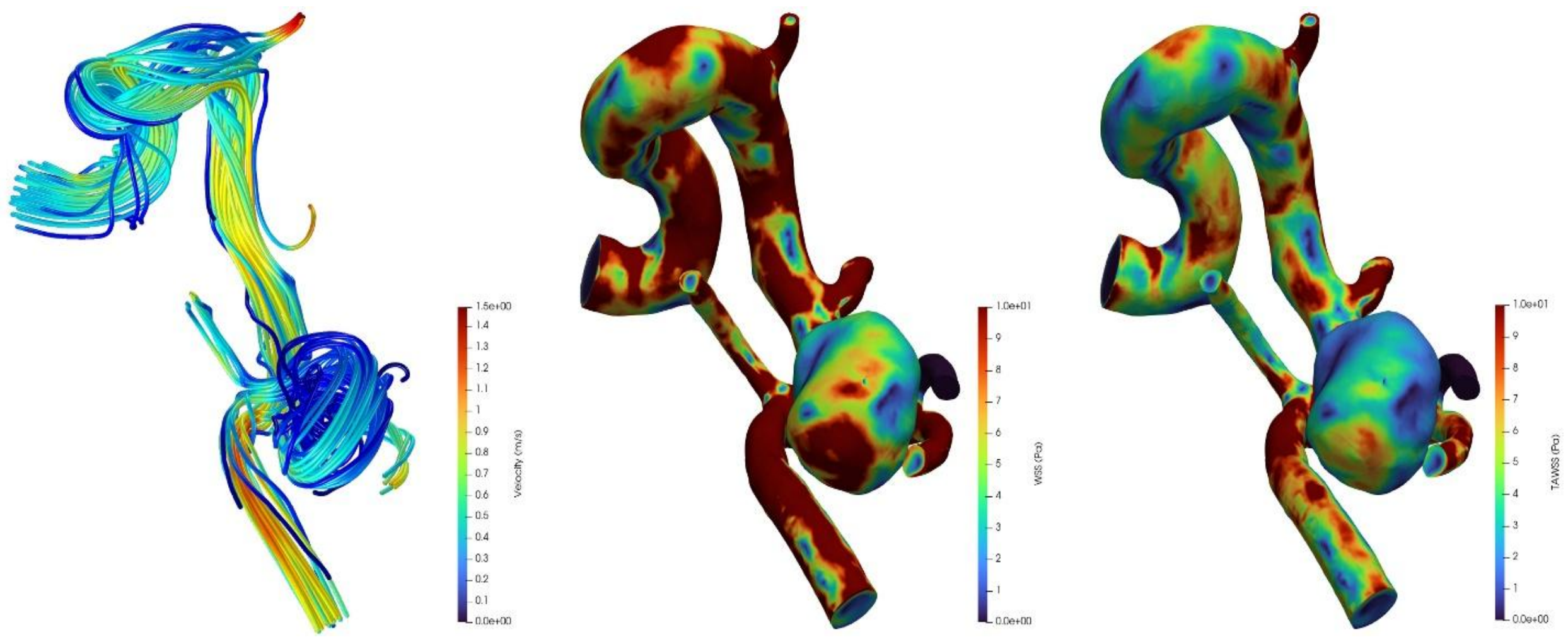

Figure 9: Velocity, WSS, and TAWSS distributions for patient C0094 (female, 36 years old, ruptured aneurysm).

## 4. Discussion

The computed velocity streamlines, WSS, and TAWSS distributions demonstrated substantial hemodynamic variability across the investigated aneurysm geometries while simultaneously revealing notable similarities between ruptured and unruptured cases. In both groups, complex intra-aneurysmal flow structures were observed, including concentrated inflow jets, recirculation zones, and swirling flow patterns within the aneurysm sacs. These flow structures produced localized regions of elevated WSS and TAWSS near aneurysm necks, bifurcation regions, and areas directly impacted by incoming flow jets.

For the unruptured aneurysms (Figures 2-5), elevated WSS and TAWSS distributions were observed in regions of strong inflow interaction and along curved vascular segments. Cases C0001 and C0020 exhibited pronounced localized shear concentrations near the aneurysm neck and adjacent parent vessels, while C0024 and C0026 demonstrated complex vortex-dominated intra-sac flow structures accompanied by heterogeneous shear distributions across the aneurysm surfaces. Despite being clinically unruptured, these cases still displayed high-magnitude WSS and TAWSS regions comparable to those typically associated with rupture-related hemodynamic environments.

Similarly, the ruptured aneurysms (Figures 6-9) exhibited complex flow behavior characterized by high-velocity inflow jets, rotational flow structures, and spatially varying wall shear distributions. Elevated WSS and TAWSS regions were observed near bifurcation zones, aneurysm necks, and localized impingement regions. However, the overall magnitude and spatial characteristics of these hemodynamic quantities showed considerable overlap with those observed in the unruptured group. In particular, cases C0068 and C0094 demonstrated extensive regions of elevated WSS and TAWSS that were qualitatively similar to patterns identified in unruptured geometries such as C0020 and C0026.

The comparative analysis indicates that elevated WSS and TAWSS are not confined exclusively to ruptured aneurysms and may also occur in stable vascular configurations. In addition, similar streamline topologies and vortex structures were identified across both groups, suggesting that complex intra-aneurysmal flow behavior alone may not uniquely distinguish rupture status. These observations highlight the multifactorial nature of cerebral aneurysm rupture and suggest that interpretation of WSS- and TAWSS-based hemodynamic analyses should be performed with caution when considered as isolated indicators of rupture risk.

Overall, the present results demonstrate substantial hemodynamic overlap between ruptured and unruptured cerebral aneurysms, emphasizing the difficulty of establishing universal rupture-specific thresholds based solely on WSS or TAWSS magnitudes.

## 5. Conclusion

Computational fluid dynamics simulations performed on eight patient-specific cerebral aneurysm models demonstrated that elevated WSS and TAWSS distributions can occur in both ruptured and unruptured aneurysms. Although the investigated geometries exhibited different anatomical configurations and flow topologies, localized high-shear regions were consistently observed near aneurysm necks, bifurcation zones, and inflow impingement regions across both groups. Similarly, complex intra-aneurysmal flow structures, including recirculation zones and vortex-dominated streamline patterns, were identified in ruptured as well as unruptured cases.

Qualitative comparison of the computed hemodynamic fields did not reveal a distinct separation between rupture states based solely on WSS or TAWSS characteristics. In several cases, unruptured aneurysms exhibited shear distributions and flow structures comparable to, or locally stronger than, those observed in ruptured geometries. These observations indicate that elevated shear-related hemodynamic environments are not unique to rupture status within the investigated dataset.

The present results therefore demonstrate that similar WSS- and TAWSS-based hemodynamic patterns may exist in aneurysms with different clinical outcomes. Within the limitations of the investigated cases and boundary condition assumptions, the study highlights the difficulty of interpreting rupture behavior using isolated shear-based parameters alone and emphasizes the highly case-dependent nature of cerebral aneurysm hemodynamics.

### Conflict of Interest

The authors declare that they have no conflict of interest.

### Data Availability

The aneurysm geometries analyzed in this study were obtained from the publicly available AneuRisk database. The code and computational data generated during the current study are available from the corresponding author upon reasonable request.

### Funding

This research received no external funding.

### Ethical Approval

This study used anonymized publicly available vascular geometries from the AneuRisk database and did not involve direct human or animal participation.

### Acknowledgments

The authors acknowledge the AneuRisk project for providing the patient-specific aneurysm geometries used in this study.

### Limitations

The present study has several limitations. Patient-specific inlet and outlet hemodynamic measurements were not available for the investigated cases; therefore, generalized physiological boundary conditions and resistance-based outlet assumptions were employed. In addition, vessel walls were modeled as rigid, and only WSS- and TAWSS-based hemodynamic quantities were investigated. The analysis was limited to eight patient-specific aneurysm models from the AneuRisk database, and the findings should therefore be interpreted as a qualitative computational comparison rather than a clinical rupture prediction study.